%% file: main.tex
\documentclass[preprint]{sig-alternate}
\usepackage{graphicx}
\usepackage{cite}
\usepackage{verbatim,color,framed}
\usepackage{subfigure}
\usepackage{amssymb,amsmath}
\usepackage{url}
\usepackage{xspace}

\long\def\com#1{}

\long\def\xxx#1{}

\def\widowpage{\pagebreak}

\renewenvironment{itemize}{
   \begin{list}{\labelitemi}{
     \setlength{\topsep}{0.5ex}
     \setlength{\itemsep}{-0pt}
     \setlength{\itemindent}{0pt}
     \setlength{\leftmargin}{\labelwidth}
     \addtolength{\leftmargin}{-8pt}}
}{\end{list}}

\newcommand{\Dissent}{Dissent\xspace}
\newcommand{\dissent}{Dissent\xspace}

\newcommand{\signed}[3]{\{#3,n_R,h_#2\}\textsc{sig}_{u_{#1}}}
\newcommand{\E}[2]{\{#1\}_{\ensuremath{#2}}}
\newcommand{\ER}[3]{\{#1\}_{\ensuremath{#2}}^{\ensuremath{#3}}}

\newcommand{\hash}[1]{\ensuremath{\textsc{hash}\{#1\}}}
\newcommand{\rand}[2]{\ensuremath{\textsc{prf}\{#1, #2\}}}

\newcommand{\allData}[1]{\ensuremath{(D_1, \ldots,D_N)_{#1}}}

\newcommand{\goflag}[1]{\ensuremath{\textsc{go}_{#1}}}

\newcommand{\true}{\textsc{true}}
\newcommand{\false}{\textsc{false}}

\definecolor{shadecolor}{rgb}{1,1,0}

\title{Accountable Anonymous Group Messaging \\
	{\tt UNPUBLISHED DRAFT}}

\numberofauthors{2}
\author{
\alignauthor
	Henry Corrigan-Gibbs \\
	\affaddr{Yale University}
\alignauthor
	Bryan Ford \\
	\affaddr{Yale University}
}

\begin{document}
\maketitle

\input{abs}
\input{intro}

\input{overview}

\input{shuffle}

\input{bulk}

\input{usage}

\input{impl}

\widowpage
\input{related}

\input{concl}


\bibliographystyle{plain}
\bibliography{main}

\com{
\pagebreak
\appendix

\input{complexity}

} 
\end{document}

%% file: abs.tex
\begin{abstract}
Users often wish to participate in online groups anonymously,
but misbehaving users may abuse this anonymity to spam or disrupt the group.
Messaging protocols such as Mix-nets and DC-nets leave online groups
vulnerable to denial-of-service and Sybil attacks,
while accountable voting protocols 
are unusable or inefficient for general anonymous messaging.

We present the first general messaging protocol
that offers provable anonymity with accountability
for moderate-size groups,
and efficiently handles unbalanced loads
where few members have much data to transmit in a given round.
The $N$ group members first cooperatively shuffle
an $N\times N$ matrix of pseudorandom seeds,
then use these seeds in $N$ ``pre-planned'' DC-nets protocol runs.
Each DC-nets run transmits the variable-length bulk data
comprising one member's message,
using the minimum number of bits required for anonymity
under our attack model.
The protocol preserves message integrity
and one-to-one correspondence between members and messages,
makes denial-of-service attacks by members traceable to the culprit,
and efficiently handles large and unbalanced message loads.
A working prototype demonstrates the protocol's practicality
for anonymous messaging in groups of $40+$ member nodes.
\com{
Each member sends $L_{tot} + O(N^2)$ bits per round,
where $L_{tot}$ is the total number of bits
all members wish to transmit in that round.
}
\end{abstract}

%% file: intro.tex
\section{Introduction}
\label{sec:intro}

\xxx{Change title?}

Anonymous participation
is often considered a basic right in free societies~\cite{yale61}.
The limited form of anonymity the that Internet provides
is a widely cherished feature~\cite{teich99,wallace99} that
enables people and groups with controversial or unpopular views
to communicate and organize without fear of personal reprisal~\cite{stein03}.
In spite of its benefits, anonymity
makes it difficult to trace or exclude
misbehaving participants~\cite{davenport02}.
Online protocols providing stronger anonymity,
such as mix-networks~\cite{chaum81untraceable,goldschlag99onion}
and DC-nets~\cite{chaum88dining,waidner89dining,sirer04eluding}
further weaken accountability and
yield forums in which no content may be considered trustworthy
and no defense is available against anonymous misbehavior.

\com{
CAPTCHAs protecting open-access forums~\cite{ahn03captcha}
lock out impaired users~\cite{chong03, may05}
and are defeatable
by artificial intelligence~\cite{chellapilla05}
or social engineering~\cite{doctorow04solving}.
Wikipedia progressively tightens its editing rules
to combat the rising tide of anonymous
vandalism~\cite{knight05, hafner06, thompson06}.
Online voting and peer review systems like Slashdot
operate reliably only to the extent that
nobody cares about the results enough
to bother opening multiple accounts
and stuffing the ballot boxes~\cite{hsieh06}.
Banning detected abusers by IP address
frequently prevents access by other legitimate users
on the same ISP~\cite{kalsey04},
and many attacks come from compromised zombie machines
not under the control of their owners~\cite{evers05}.
}

This paper focuses on providing anonymous messaging
within small, private online groups.
We assume a group's membership is closed and known to its members;
creating groups with secret membership
is a related but orthogonal goal~\cite{vasserman09membership}.
Members may wish to send messages to each other,
to the whole group, or to a non-member,
such that the receiver knows that {\em some} member sent the message
but no one knows {\em which} member.
\com{
assuring the receiver that {\em some} member sent the message,
but giving neither the receiver, nor other group members,
nor a powerful adversary
who has compromised some members and monitors all network traffic,
any information about {\em which} member sent the message.
}
Members may also wish to cast secret ballots in votes held by the group,
or to create pseudonyms under which to collaborate with other members.

We also wish to hold members {\em accountable}, however:
{\em not} by compromising their anonymity
and allowing some authority or majority quorum
to unmask a member whose messages prove unpopular,
but rather by ensuring that no malicious member
can abuse his (strong) anonymity to disrupt the group's operation.
For example,
a malicious member should be unable to
corrupt or block other members' messages,
overrun the group with spam,
stuff ballots,
or create unlimited anonymous Sybil identities~\cite{douceur02sybil}
or sock puppets~\cite{stone07}
with which to bias or subvert the group's deliberations.

\com{
This paper develops
a group communication primitive we call {\em shuffled send}.
In a shuffled send communication round,
each member of a group anonymously submits
exactly one message,
all of which are randomly shuffled
and revealed to a designated target.
The target may be a particular group member,
a non-member requesting a service from the group,
or all group members;
the last case provides receiver as well as sender anonymity.
\xxx{Do we provide receiver anonymity?}
Shuffled send provides anonymity within a well-defined group
whose membership is known at least to its members;
creating groups with secret membership
is a related but orthogonal goal~\cite{vasserman09membership}.
Shuffled send may be useful in building many online applications
such as anonymous messaging forums,
voting and deliberative processes,
data mining~\cite{DBLP:conf/kdd/BrickellS06},
file sharing~\cite{sirer04eluding},
collaborative editing~\cite{stone07},
and auctions~\cite{stajano99cocaine}.
}

\com{
While secret shuffles are well-studied~\cite{
	chaum81untraceable,neff2001voting,furukawa01efficient},
we wish to provide two properties
that existing shuffle protocols generally do not:
{\em accountability}, 
enabling the group to trace and exclude any member
maliciously disrupting the protocol's operation,
without compromising the anonymity
of any member correctly following the protocol;
and {\em communication efficiency}
even under unbalanced message loads,
where one member wishes to send a large message
while others have nothing to send.
Strong anonymity protocols
such as DC-nets~\cite{chaum88dining,waidner89dining}
are difficult to protect against anonymous disruption by malicious members,
and provide no protection against Sybil attacks~\cite{douceur02sybil},
such as online ballot stuffing or sock puppets~\cite{stone07}.
Both DC-nets and shuffle schemes~\cite{
	chaum81untraceable,neff2001voting,furukawa01efficient,
	DBLP:conf/kdd/BrickellS06}
generally require all members
to send equal-size messages or transmit at a similar rate,
wasting $O(N)$ bandwidth when only one member has data to send.

Online groups could use mix-networks~\cite{
	chaum81untraceable,goldschlag99onion}...
}

As a motivating example,
suppose an international group of journalists
wishes to form a ``whistleblowing'' publication
like WikiLeaks~\cite{wikileaks}.
To protect journalists and their sources
more strongly than the world's varied legal frameworks do,
member journalists wish to submit leaked documents and related information
to the group anonymously.
Members need
assurance that powerful organizations or governments
cannot trace the leak to an individual journalist or her source.
The journalists wish to prove to their readers
that leaked documents come via a trustworthy channel,
namely one of the group's known and reputable members,
and not from an outsider.
The group must be able to analyze and vet each document thoroughly
before collectively approving it for publication.
The group must protect its internal operation and its members' anonymity
even from adversaries who have planted colluding spies within the group.
\com{
Since powerful adversaries may control some members,
the group must protect members' anonymity from colluding spies,
the group's internal votes must be protected from ballot-stuffing,
internal deliberation must be protected from bias via sock puppetry,
and the group must function even when spies
attempt to disrupt communication.
Other applications for accountable anonymous group messaging include
data mining~\cite{DBLP:conf/kdd/BrickellS06},
file sharing~\cite{sirer04eluding},
collaborative editing~\cite{stone07},
and auctions~\cite{stajano99cocaine}.
}
And this security must come at acceptable time and resource costs.
\com{	XXX say this below...
Further, leaked documents may be megabytes or gigabytes in size,
and often only one member will have a document to leak at a given time,
creating unbalanced loads
that current strong anonymity protocols handle inefficiently.
}

We present an accountable anonymous messaging protocol called \dissent
(Dining-cryptographers Shuffled-Send Network),
the first we know of
with the properties needed in scenarios like the one above.
\Dissent provides provable integrity, anonymity, and accountability
in the face of strong traffic analysis and compromised members,
and an experimental prototype shows it to be efficient enough
for latency-tolerant messaging in small but widely distributed groups.
\com{
\Dissent ensures anonymity and message integrity
even if the entire network and many group members
are under an attacker's control.
The protocol enforces a one-to-one correspondence
between group members
and messages resulting from a given run,
making it usable as a group voting protocol
resistant to Sybil attacks~\cite{douceur02sybil}
from participating group members.

\Dissent aims to provide anonymous communication within
a \em{closed} group of participants.  Since group members
are well-known, we assume that a single attacker
cannot obtain multiple memberships within the
group.  We protect against a Sybil ``ballot-stuffing''
attack, where one group member attempts to submit many
messages in a single round of the protocol.

\Dissent is designed for moderate-size groups,
containing tens or perhaps a hundred direct participants,
and is not intended for ``open-access''
anonymous messaging or file sharing~\cite{
	goldschlag99onion,clarke00freenet},
though it may be a useful component
in designs such as Herbivore~\cite{sirer04eluding}.
\Dissent is similarly not intended as a large-scale election system,
though in the context of small groups
it can provide vote integrity, anonymity, and limited coercion resistance.
}

In contrast with mix-networks~\cite{chaum81untraceable,goldschlag99onion}
and DC-nets~\cite{chaum88dining,waidner89dining,sirer04eluding},
\dissent implements a {\em shuffled send} primitive,
where each group member sends
{\em exactly} one message per round,
making it usable for voting or assigning pseudonyms
with a 1-to-1 correspondence to real group members.
Unlike verifiable cryptographic shuffles~\cite{
	neff2001voting,furukawa01efficient},
\dissent uses only readily-available cryptographic primitives,
and handles arbitrary-size messages and unbalanced loads efficiently,
such as when one journalist has a multi-gigabyte document to leak
at a time when the others have nothing to send.

\com{
\Dissent provides {\em accountability}
in the sense that a group can trace and identify any member
attempting to disrupt the group's operation
via denial-of-service (DoS) attacks,
and members cannot flood the group with fake pseudonyms or stuff ballots.
Our notion of accountability does {\em not} include
enabling some threshold or majority of members
to unmask an otherwise well-behaved member
whose message content or voting record is unpopular:
\dissent protects a member's anonymity among all honest members,
even in the presence of any number of maliciously colluding members.
}

\com{
Our shuffled-send primitive has potential
applications to secret-ballot elections, but the scheme
as we present here is not suitable for a general-purpose
election protocol.  Even ignoring issues of scalability, 
our shuffled-send primitive does not have the property of
coercion resistance.  We discuss possible ways to achieve
coercion resistance in section \ref{sec-usage}.
}

\xxx{from Jacob: I think I'd like to see a different set of figures.  Maybe one near the intro with what the components are going to be, and how they interact (the current Figure 1 doesn't quite do this).  Then one on the details of the shuffled-send protocol (to go along with section 2.4?), and finally one like the existing figure 2, but see if you can put it earlier in section 4.}

\Dissent operates in two stages,
{\em shuffle} and {\em bulk transfer}.
The shuffle protocol builds on a data mining
protocol by Brickell and Shmatikov~\cite{DBLP:conf/kdd/BrickellS06}
to permute a set of fixed-length messages,
one from each group member,
and broadcast the set to all members
with cryptographically strong anonymity.
Like many anonymous messaging protocols,
the original data mining protocol was vulnerable
to untraceable denial-of-service (DoS) attacks by malicious members.
Our refinements remove this vulnerability
by adding {\em go/no-go} and {\em blame} phases,
which can trace and hold accountable
any group member maliciously disrupting the protocol.

\Dissent's bulk protocol builds on
the information-theoretic anonymity of DC-nets~\cite{
	chaum88dining,waidner89dining,sirer04eluding},
but leverages \dissent's shuffle protocol
to replace the DoS-prone slot reservation systems in prior DC-nets schemes
with a prearranged transmission schedule
guaranteeing each member exactly one message slot per round.
In each round,
all group members broadcast bit streams
based on pseudorandom seeds distributed via the shuffle protocol,
so that XORing all members' bit streams together
yields a permuted concatenation of all members' variable-length messages.
\com{
Also unlike prior protocols,
\dissent does not require members to pad messages
or reserve equal bandwidth shares:
if one member has $L$ bits to send and all others have none,
then each member sends only $L+O(1)$ bits in the bulk phase,
the minimum possible without compromising anonymity under traffic analysis.
}
Cryptographic hashes distributed in the shuffle phase
enable members to verify the correctness of each others' bulk transmissions,
ensuring message integrity and DoS protection throughout.

\xxx{from Jacob: the no padding point a bit earlier: The case I think you want to point out is not when there's a single sender, but when there are multiple senders.  I think I'm a bit confused about what you count as padding?  If all members need to transmit $L+O(1)$ bits but most of the messages are much smaller than L, what do you call the rest.  (I think when you get to this later in the paper it's clear, but at that point I'm not sure }

\xxx{ make sure it's stated clearly exactly why anonymity with accountability
	is challenging.}

\xxx{summarize implementation and experimental results.}

\Dissent has limitations, of course.
It is not intended for large-scale, ``open-access''
anonymous messaging or file sharing~\cite{
	goldschlag99onion,clarke00freenet},
although it might serve as a building block
in designs like Herbivore~\cite{sirer04eluding}.
\Dissent's accountability properties assume closed groups,
and are ineffective if a malicious member can just leave and rejoin the group
under a new (public) identity after expulsion.
\Dissent is also not a general-purpose voting system,
providing only a limited form of coercion resistance for example.
The serialized shuffle protocol imposes a per-round startup delay
that makes \dissent impractical for latency-sensitive applications.

We built a working prototype of \dissent,
and tested it under Emulab~\cite{emulab} on groups of up to 44 nodes
connected via simulated wide-area links.
Anonymously distributing messages up to 16MB in size
among 16 nodes with 100ms inter-node delays,
\dissent's shuffle and other startup costs incur a 1.4-minute latency,
but it handles large message loads, both balanced and unbalanced,
in about $3.5\times$ the time required
for non-anonymized group messaging via TCP.
Varying group size,
\dissent can send a 1MB message anonymously
in less than 1 minute in a 4-member group,
4 minutes for a 20-node group,
and 14 minutes for a 40-node group.
While not suitable for interactive workloads, therefore,
\dissent should be usable
for ``WikiLeaks''-type scenarios requiring strong security guarantees
in small decentralized groups.

This paper makes four main technical contributions.
First, we enhance Brickell/Shmatikov's shuffle protocol~\cite{
	DBLP:conf/kdd/BrickellS06}
to make DoS attackers traceable without compromising anonymity.
Second, we use this shuffle protocol to create
a DoS-resistant DC-nets variant for bulk transfer,
which guarantees each member exactly one transmission slot per round.
Third, we introduce the first shuffle protocol
that supports arbitrary-size and unbalanced message loads efficiently,
e.g., when only one member has data to send.
Fourth, we demonstrate through a working prototype
the practicality of the protocol,
at least for delay-tolerant applications.

Section~\ref{sec-overview} provides an overview
of \dissent's communication model, security goals, and operation.
Section~\ref{sec-shuffle} formally describes the shuffle protocol, and
Section~\ref{sec-bulk} details the bulk protocol.
Section~\ref{sec-usage} informally covers
practical implementation and usage considerations
such as protocol initiation, coercion resistance, and liveness.
Section~\ref{sec-impl} describes our prototype implementation
and experimental results.
Section~\ref{sec-related} summarizes related work, and
Section~\ref{sec-concl} concludes.

%% file: overview.tex
\section{Protocol Overview}
\label{sec-overview}

This section first introduces the group communication model
our protocol implements,
outlines a few applications of this model,
and defines the protocol's precise security goals,
leaving protocol details to subsequent sections.

\subsection{The Shuffled Send Primitive}

The purpose of \dissent is to provide a {\em shuffled send}
communication primitive,
providing sender anonymity among a well-defined group of nodes.
We assume that the set of members comprising the group,
and each member's public key or certificate,
is agreed upon and known to all group members.
The group may initiate a run of the shuffled send protocol
in any way that preserves anonymity goals:
e.g., a designated leader, or {\em every} group member,
might initiate runs periodically,
or a ``client'' in or outside the group not requiring anonymity
might initiate a run to request a service provided by the group collectively.
(A member's desire to send anonymously
must not be the initiation event,
if traffic analysis protection is desired.)
Each protocol run is independent and permits each group member
to send exactly one variable-length message
to some target designated for that run;
ongoing interaction requires multiple protocol runs.
A run's target may be a particular group member,
all members (for anonymous group multicast),
or another node such as a non-member ``client'' that initiated the run.

\begin{figure}[t]
\centering
\includegraphics[width=0.45\textwidth]{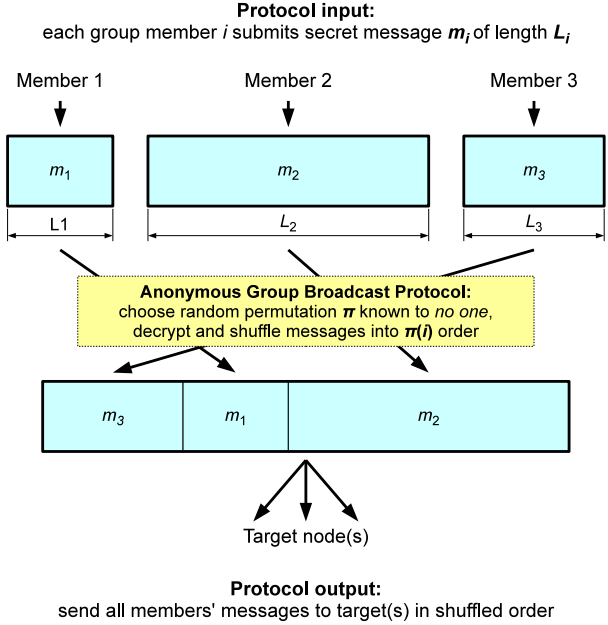}
\caption{Shuffled send communication model}
\label{fig-model}
\end{figure}

Each protocol run operates
as shown in Figure~\ref{fig-model}.
Every group member $i$ secretly creates a message $m_i$
and submits it to the protocol.
The protocol collects all $N$ secret messages,
shuffles their order according to some random permutation $\pi$
that {\em no one} knows,
concatenates the messages in this shuffled order
so that $m_i$ appears at position $\pi_i$,
and sends the concatenated sequence of messages
to the target.
Each input message $m_i$ can have a different length $L_i$,
and the protocol's output has length $\sum_i L_i$.

\com{
A variety of events could trigger a run of the protocol.
For example, any member (or even a non-member)
might unilaterally initiate a run of the protocol,
if the initiator wishes to obtain some service from the group
acting as a collective:
in this case, the initiator's action is not anonymous,
although the group members' responses will be.
If the group wishes to remain in regular contact
without relying on such initiation events,
they might form a regular schedule,
or members might initiate runs at random intervals...
Or by consensus...
}

\subsection{Applications of Shuffled Send}

The shuffled send model
combines and generalizes the functionality
of several classes of anonymity protocols.
Although every participant must submit a message
in a given protocol run,
members with nothing to send can submit a message of length zero,
providing efficient single-sender as well as multiple-sender service.
(The protocol still causes each member to send a similar number of bits
on the underlying network for traffic analysis protection,
but none of these bits are wasted padding messages of unbalanced lengths.)
Members wishing receiver anonymity
can first anonymously send a public encryption key to establish a pseudonym,
then look for messages encrypted with that key
in subsequent shuffled sends targeted at the whole group.

Since each member may submit {\em exactly} one message per shuffled send,
one run's messages can serve as ballots in an anonymous vote.
Unlike anonymous voting protocols
designed for specific types of ballots and tallying methods,
\dissent supports ballots of arbitrary type, format, and size,
and group members can count and independently verify the ballots
in any agreed-upon fashion.  
Ballots need not be one-shot messages either:
a group can use one protocol run to establish
a set of pseudonymous signing keys, exactly one per member,
then use these pseudonyms in subsequent protocol runs
for pseudonymous deliberation,
without permitting members to create unlimited pseudonyms
for Sybil attacks~\cite{douceur02sybil}
or sock puppetry~\cite{stone07}.  

\com{
provided by both conventional anonymous messaging protocols,
which typically treat individual messages independently,
and voting protocols,
which treat ``messages'' (ballots) as a group,
usually to be tallied in some predefined fashion.
Anonymous group broadcast easily implements individual message broadcast:
if only one member has anything to send in a certain protocol run,
all other members can simply send empty messages.
Similarly, anonymous group broadcast trivially allows
all manner of voting schemes to be implemented,
since the one-to-one relationship between secret inputs and shuffled outputs
ensures that the resulting vote count may be verified by all group members,
whatever the precise format of ballots and method of counting them.
This combination provides richer functionality
than either of the conventional communication models it builds on,
however...
}

Applications for which shuffled send may be suited
include whistleblowing~\cite{wikileaks},
surveys~\cite{DBLP:conf/kdd/BrickellS06},
file sharing~\cite{sirer04eluding},
accountable Wiki-style editing~\cite{stone07},
and ``cocaine auctions''~\cite{stajano99cocaine}.
The current version of \dissent also has limitations:
e.g., it may not scale to large groups,
it provides only a limited form of coercion resistance
described in Section~\ref{sec-coercion},
and the latency of the shuffle required on each protocol run
may make the protocol impractical for interactive or real-time messaging.
Future work may be able to address these limitations.


\input{goals}

\subsection{Protocol Operation Summary}

\Dissent consists of two sub-protocols:
a {\em shuffle} protocol and a {\em bulk} protocol,
whose operation we briefly summarize here
to provide context for the detailed descriptions
in the next sections.

In the shuffle protocol,
all members $1,\dots,N$ first choose secret messages $m_1,\dots,m_N$,
of {\em equal} length $L$.
Each member $i$ now iteratively wraps its message $m_i$
in $2N$ layers of public-key encryption using an 
IND-CCA2 \cite{bellare1998cryptodefs}
secure algorithm.
Member $i$ first encrypts $m_i$
using a list of temporary {\em secondary public keys} $z_j$,
one for each member $j$, in reverse order $z_N,\dots,z_1$,
to yield an intermediate cipherext $C'_i$.
Member $i$ then encrypts $C'_i$ further
using a list of {\em primary public keys} $y_N,\dots,y_1$
to form a final ciphertext $C_i$.

Member $1$ collects all final ciphertexts into one list,
then each member $i$ in turn takes this list,
strips off one layer of encryption using his primary private key $x_i$,
randomly shuffles the list, and passes the result to $m_{i+1}$.
Member $N$ broadcasts the final shuffled list to all members,
each of whom verifies that the list includes
her own intermediate ciphertext $C'_i$,
and broadcasts a {\em go} if so and a {\em no-go} otherwise.

Each member $i$, upon receiving a {\em go} from {\em all} members,
broadcasts her secondary private key $w_i$ associated with $z_i$,
enabling all members to decrypt the shuffled messages.
On receiving a {\em no-go} from {\em any} member, however,
member $i$ destroys her private key $w_i$ and enters a {\em blame} phase,
where all members reveal the secrets used
to encrypt the intermediate ciphertexts.
Our shuffle protocol ensures integrity and anonymity
exactly as in its precursor~\cite{DBLP:conf/kdd/BrickellS06},
but our new {\em go/no-go} and {\em blame} phases
enable all group members to trace the culprit of any protocol malfunction.

The shuffle protocol has two practical limitations:
all messages must be of equal length $L$,
incurring $O(NL)$ extra communication
if only one member wishes to send;
and its decrypt-and-shuffle phase is inherently serial,
incurring a long delay if $N$ or $L$ is large.
We currently have no solution if $N$ is large,
but our {\em bulk} protocol addresses the problem
of sending large, variable-length messages efficiently.

\begin{figure*}[t]
\centering
\includegraphics[width=0.75\textwidth]{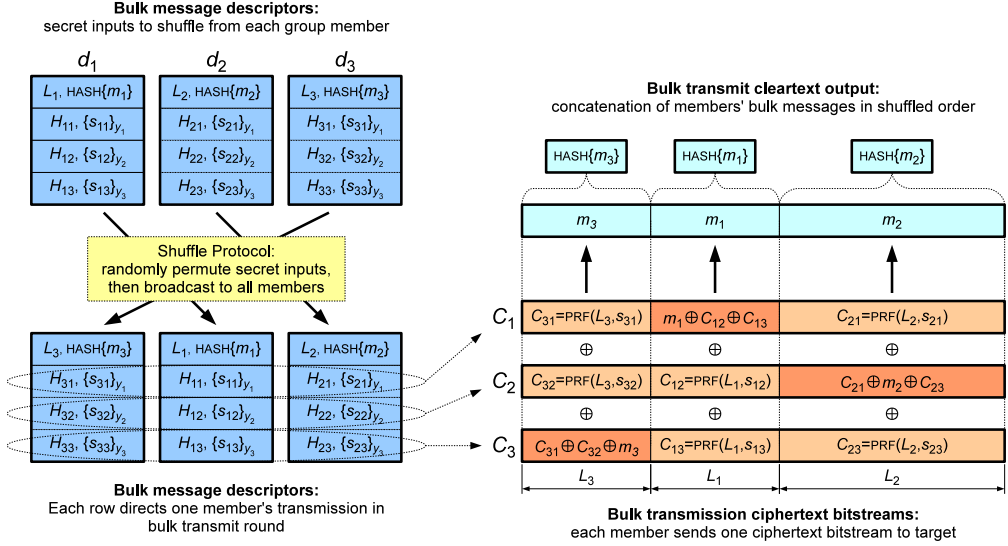}
\caption{Illustration of bulk protocol operation for 3-member group,
	shuffled using permutation $\pi = [2,3,1]$.}
\label{fig-bulkproto}
\end{figure*}

As illustrated in Figure~\ref{fig-bulkproto},
the bulk protocol uses the shuffle protocol
to shuffle a set of $N$ {\em message descriptors},
one submitted anonymously by one member, instead
of shuffling the messages themselves.
Each descriptor $d_i$ contains the length $L_i$
of member $i$'s message $m_i$,
a cryptographic hash of $m_i$,
a vector $\vec{S}_i$ of $N$ seeds $s_{ij}$,
each seed encrypted with $j$'s primary public key
and assigning $j$ a pseudo-random bulk ciphertext to transmit,
and a vector $\vec{H}_i$ of hashes $H_{ij}$
validating each bulk ciphertext.

Member $i$ ``assigns himself'' a junk seed $s_{ii}$
and a hash $H_{ii}$ of a ciphertext that,
when XORed with the ciphertexts $i$ ``assigned'' other members,
yields $i$'s message $m_i$.
Once the shuffle protocol has revealed
the $N$ shuffled message descriptors,
representing an $N \times N$ matrix of bulk ciphertext ``assignments,''
group members send (in parallel) their assigned ciphertexts
to the designated target,
enabling the target to recover and verify all members' messages.
If any member produces an incorrect bulk ciphertext,
a {\em blame} phase reruns the shuffle protocol,
enabling the anonymous sender of the corrupted message
to ``accuse'' and expose the culprit.

\subsection{Simplifying Assumptions}

Our core protocol descriptions in Sections~\ref{sec-shuffle} and \ref{sec-bulk}
make several simplifying assumptions,
which we will relax and address more realistically
later in Section~\ref{sec-usage}.
We assume for now that:
(a) all members know when to initiate a protocol run
	and how to distinguish one run from another;
(b) all members of a group participate in every protocol run;
(c) all members have public encryption keys
	and nonrepudiable signing keys known to all other members;
	and
(d) all members remain connected throughout a protocol run
	and never stop sending correctly-signed messages,
	until the protocol run has completed
	from the perspective of all group members.
Assumption (d) implies that we address only safety properties for now,
deferring liveness issues to Section~\ref{sec-usage}---%
including the important corner case of
a node withholding the last message it is supposed to send
while collecting all other members' final messages,
learning a protocol run's results
while denying others those results.

%% file: goals.tex
\subsection{Security Goals}
\label{sec-goals}

We now precisely define \dissent's attack model and security goals.
We assume the attacker is polynomial-time limited,
but can monitor all network traffic
and compromise any subset of group members.
A member is \textit{honest} if she follows the protocol exactly
and is not under the attacker's control,
and {\em faulty} otherwise.
Faulty nodes may collude and send arbitrary messages.
For simplicity, our core protocol descriptions
in Sections~\ref{sec-shuffle} and \ref{sec-bulk}
assume that nodes never just go silent;
we address liveness using principles from PeerReview~\cite{
	haeberlen07peerreview}
as outlined in Section~\ref{sec-usage}.
\com{
Participants who stop sending messages during the protocol's
execution may be doing so honestly (if the network connection
failed, for example) or with the intention of disrupting
the protocol (dishonestly).  The problem of how to distinguish
between these two types of ``silent" participants is outside
of the scope of this paper.  We refer interested readers to
PeerReview for further discussion of this topic
\cite{haeberlen07peerreview}.
}
The formal security properties we wish the protocol to satisfy
are {\em integrity}, {\em anonymity}, and {\em accountability},
as we define below.

\begin{itemize}
\item	{\bf Integrity:}
	The protocol maintains \textit{integrity}
	if at the end of a protocol run,
	every honest member either:
	(a)	obtains exactly $N$ messages,
		including exactly one submitted by each honest member, or
	(b)	knows that the protocol did not complete successfully.

\item	{\bf Anonymity:}
	Following Brickell and Shmatikov~\cite{
		DBLP:conf/kdd/BrickellS06},
	the protocol maintains {\em anonymity}
	if a group of $k \leq N-2$ colluding members
	cannot match an honest participant's message to its author
	with a probability significantly better than random guessing.
	(If all but one member colludes, no anonymity is possible.)

\item	{\bf Accountability:}
	Adopting ideas from PeerReview~\cite{haeberlen07peerreview},
	a member $i$ {\em exposes} a member $j$
	if $i$ holds third-party verifiable proof of $j$'s misbehavior.
	The protocol maintains \textit{accountability} if
	no member ever exposes an honest member,
	and after a run, either:
	(a)	each honest member successfully obtains
		every honest member's message, or
	(b) 	all honest members expose at least one faulty member.
\end{itemize}


%% file: shuffle.tex
\section{Shuffle Protocol}
\label{Shuffle}
\label{sec-shuffle}

This section details the shuffle protocol,
first covering its cryptographic building blocks,
then formally describing protocol,
proving its correctness,
and analyzing its complexity.


\subsection{Cryptographic Primitives}
\label{sec-shuffle-prim}

We use a standard, possibly randomized \textit{signature scheme}
consisting of:
(a) a key generation algorithm producing a private/public key pair $(u,v)$;
(b) a signing algorithm taking private key $u$ and message $m$
	to produce signature $\sigma = \textsc{sig}_u\{m\}$; and
(c) a deterministic verification algorithm taking public key $v$, message $m$,
	and candidate signature $\sigma$, and returning true
	iff $\sigma$ is a correct signature of $m$
	using $v$'s associated private key $u$.
The notation $\{m\}\textsc{sig}_{u}$ indicates the concatenation
of message $m$ with the signature $\textsc{sig}_u\{m\}$.

We also require a {\em public-key cryptosystem},
which must be IND-CCA2 secure~\cite{bellare1998cryptodefs}
(e.g., RSA-OAEP~\cite{Fujisaki:2004fk}).
The cryptosystem consists of:
(a) a key generation algorithm producing a private/public key pair $(x,y)$;
(b) an encryption algorithm taking public key $y$, plaintext $m$,
and some random bits $R$, and producing a ciphertext $C = \{m\}_y^R$;
(c) a deterministic decryption algorithm taking private key $x$
	and ciphertext $C$, and returning the plaintext $m$.
We assume a node can save
the random bits $R$ it uses during encryption,
and that it can encrypt deterministically using a given $R$,
such that given inputs $y$, $m$, and $R$
always yield the same ciphertext.
Software cryptosystems using pseudorandom number generators
generally satisfy this assumption.
The notation $C = \{m\}_{y_1:y_N}^{R_1:R_N}$ indicates
iterated encryption via multiple keys:
$C = \{\dots\{m\}_{y_1}^{R_1}\dots\}_{y_N}^{R_N}$.
We omit $R$ when an encryption's random inputs need not be saved.


\com{
We adopt the definition of a \textit{public-key cryptosystem} from
Fujisaki et al. \cite{Fujisaki:2004fk}.
We modify this definition slightly, requiring that the
randomness $r$ used to encrypt a message is recoverable
during decryption.  We also require that the cryptosystem
in use be IND-CCA2 secure \cite{bellare1998cryptodefs}.
As in Justin Brickell and Vitaly Shmatikov's paper on
anonymous data collection, we denote the encryption of
message $m$ with public key $y$ as $\E{m}{y}$
\cite{DBLP:conf/kdd/BrickellS06}.
To denote the serial encryption of message $m$ with
public keys $y_N, \ldots, y_1$, we will write
$\E{m}{y_{N}:y_1}$.
}

We use the standard definition \cite{Stinson} of an \textit{unkeyed
hash function} and will denote the hash of message $m$ as
$\hash{m}$.

We use a standard definition \cite{Stinson} of a \textit{pseudo-random bit
generator}.  We will denote the first $L$ bits generated from
a pseudo-random bit generator seeded with $s$ as \rand{L}{s}.

\subsection{Formal Protocol Description}
\label{dataexchange}

\newcounter{PhaseCounter}[subsection]
\newcommand{\phaseinner}[1]{Phase \arabic{PhaseCounter}\label{#1}}
\newcommand{\phase}[1]{\refstepcounter{PhaseCounter}\phaseinner{#1}:}
\newcommand{\subphase}[2]{Phase \arabic{PhaseCounter}#1:\label{#2}}

Each member $i$ (for $i = 1, \ldots, N$) has
a primary encryption key pair $(x_i, y_i)$,
a signing key pair $(u_i, v_i)$,
and a secret datum $d_i$ of fixed length $L$ to send anonymously.

Before a protocol run,
all members agree on
a session nonce $n_R$ uniquely identifying this protocol run,
the participants' primary public encryption and signing keys,
and a common ordering of all members $1,\dots,N$.
Such agreement might be achieved via Paxos~\cite{Lamport:1998}
or BFT~\cite{castro99practical}.

The shuffle protocol operates in {\em phases};
each member $i$ sends at most one message $m_{i\phi}$ per phase $\phi$,
though $i$ may broadcast the same $m_{i\phi}$ to several members.
Each member maintains a {\em tamper-evident log}
of all messages it sends and receives in a protocol run~\cite{
	haeberlen07peerreview}.
Member $i$ signs each $m_{i\phi}$ it sends with its private key $u_i$,
and includes in each message the session nonce $n_R$
and a hash $h_\phi$ of $i$'s current log head in phase $\phi$.
Each $h_\phi$ depends on all messages $i$ received up to phase $\phi$
before sending $m_{i\phi}$.
Members ignore any messages they receive
containing a bad signature or session nonce.

\begin{itemize}
	\item \phase{Shuffle:keypairGen} Secondary Key Pair Generation.
		Each member $i$ chooses an encryption key pair $(w_i, z_i)$,
		and broadcasts:
			\begin{center}
				$\signed{i}{1}{z_i}$
			\end{center}
	\item \phase{Shuffle:dataSubmission} Data submission.
		Each member $i$ encrypts her datum $d_i$ with all members'
			secondary public keys:	
			\begin{center}
				$C'_i = \E{d_i}{z_N:z_1}$
			\end{center}
		Member $i$ stores $C'_i$ for later use,
		then further encrypts $C'_i$ with all members'
		primary public keys,
		this time internally saving the random bits used
		in each encryption: 
			\begin{center}
				$C_i = \ER{C'_i}{y_N:y_1}{R_{iN}:R_{i1}}$
			\end{center}
		Member $i$ now sends to member 1:
			\begin{center}
				$\signed{i}{2}{C_i}$
			\end{center}

	\item \phase{Shuffle:anonymization} Anonymization.
		Member 1 collects all ciphertexts
		into a vector $\vec{C}_0 = C_1,\dots,C_N$,
		randomly permutes its elements,
		then strips one layer of encryption from each ciphertext
		using private key $x_1$ to form $\vec{C}_1$.
		Member 1 sends to member 2:
			\begin{center}
				$\signed{i}{3}{\vec{C}_1}$
			\end{center}
		Each member $1<i<N$ in turn accepts $\vec{C}_{i-1}$,
		permutes it, strips one encryption layer
		to form $\vec{C}_i$,
		then sends $\vec{C}_i$ to member $i+1$.
		Member $N$ finally permutes and decrypts $\vec{C}_{N-1}$
		to form $\vec{C}_N$,
		and broadcasts to all members:
			\begin{center}
				$\signed{i}{3}{\vec{C}_N}$
			\end{center}
		If any member $i$ detects a duplicate or invalid ciphertext
		during this phase, member $i$ reports it
		and the group moves directly to phase 5b below (``blame'').

	\item \phase{Shuffle:verification} Verification.
		All members now hold $\vec{C}_N$,
		which should be a permutation of $C'_1, \ldots, C'_N$.
		Each member $i$ verifies that her own $C'_i$
		is included in $\vec{C}_N$,
		sets a flag $\goflag{i}$ to $\true$ if so
		and $\false$ otherwise,
		and broadcasts:
			\begin{center}
				$\signed{i}{4}{\goflag{i}, \hash{\vec{C}_N}}$
			\end{center}

		Each member $i$ then waits for such a ``go/no-go'' message
		from {\em all} other members.
		If \textit{every} member $j$ reports $\goflag{j} = \true$
		for the correct $\hash{\vec{C}_N}$,
		then member $i$ enters phase 5a below;
		otherwise $i$ enters phase 5b.

	\refstepcounter{PhaseCounter} 

	\item \subphase{a}{Shuffle:decryption} Decryption.
		Each member $i$ destroys her copy of $C'_i$
		and the random bits she saved in phase 2,
		then broadcasts her secondary private key $w_i$
		to all members:
			\begin{center}
				$\signed{i}{5}{w_i}$
			\end{center}

		Upon receiving all keys $w_1, \ldots, w_N$,
		member $i$ checks that each $w_j$ is the private key
		corresponding to public key $z_j$,
		going to phase 5b if not.
		Member $i$ then removes the remaining $N$ levels of encryption
		from $\vec{C}_N$,
		resulting in a permutation of the submitted data
		$d_1, \ldots, d_N$.

	\item \subphase{b}{Shuffle:blame} Blame.
		Each member first destroys her secondary private key $w_i$,
		then reveals to all members
		the random bits $R_{ij}$ she saved
		from the primary public key encryptions in phase 2,
		and all signed messages she received and sent in phases 1--4.
		Each member $i$ uses this information
		to check the behavior of each member $j$ in phases 1--4,
		replaying $j$'s primary key encryptions in phase 2
		and verifying that $j$'s anonymized output $\vec{C}_j$
		in phase 3
		was a decrypted permutation of $\vec{C}_{j-1}$.
		Member $i$ {\em exposes} member $j$ as faulty if $j$ signed
		an invalid $z_j$ in phase 1,
		an incorrectly encrypted $C_j$ in phase 2,
		an improperly decrypted or permuted $\vec{C}_j$ in phase 3,
		a $\goflag{j} = \false$
		or a wrong $\hash{\vec{C}_N}$ in phase 4
		after phases 1--3 succeeded,
		an incorrect $w_j$ in phase 5a;
		or if $j$ equivocated by signing
		more than one message or log head $h_\phi$ in any phase $\phi$.
\end{itemize}

\subsection{Proofs of Correctness}

The shuffle protocol's integrity and anonymity
derive almost directly from Brickell/Shmatikov~\cite{
	DBLP:conf/kdd/BrickellS06},
so we only sketch proofs of these properties,
focusing instead on the accountability property
introduced by our enhancements.

\subsubsection{Integrity}
\label{Shuffle:proof:integrity}

To preserve integrity,
after a protocol run every honest member must either:
(a) hold the datum $d_i$ of every honest member $i$, or
(b) know that the protocol did not complete successfully.
Suppose that a protocol run appears to complete successfully
via phase 5a (decryption),
but some honest member $i$ does not hold the plaintext $d_j$
of some other honest member $j$.
Since $j$ is honest,
$j$'s intermediate ciphertext $C'_j$ must be a correct encryption of $d_j$,
and $C'_j$ must have appeared in $\vec{C}_N$,
otherwise $j$ would have sent $\goflag{j} = \false$ in phase 4.
Since honest member $i$ would not enter phase 5a
without receiving $\goflag{j} = \true$
for the same $\vec{C}_N$ from {\em all} members,
member $i$ must hold $C'_j$,
and $C'_j$ must decrypt to $d_j$
if all members released correct secondary private keys $w_1,\dots,w_N$
during phase 5a.
If some faulty member released an incorrect key $w'_k \ne w_k$,
all honest members see that $w'_k$ does not match
$k$'s public key $z_k$
and know that $k$ is faulty.

\com{
By way of contradiction, assume that at the end of a run of the protocol, there exists an honest member $h_{\alpha}$ who does not hold the plaintext of some other honest member and that $h_{\alpha}$ does \textit{not} know that the plaintext of an honest member $h_{\beta}$ has been modified during the protocol's execution.

Since member $h_{\beta}$ is honest by assumption, in the verification phase, member $h_{\beta}$ will notice that her ciphertext $C'_{h_{\beta}}$ is not in the collection of ciphertexts $\allData{N}$.  Member $h_{\beta}$ will send a ``No-go'' message to all other members.  Upon receipt of the ``No-go'' message, honest member $h_{\alpha}$ will know that at least the plaintext of honest member $h_{\beta}$ was modified.  This is a contradiction.

Alternatively, a dishonest member $d$ could attempt to corrupt the decrypted ciphertext of honest member $h_{\beta}$ by releasing a secondary private key $w_d$ that does not correspond to her secondary public key $z_d$.  Honest member $h_{\alpha}$ will detect this attack by noticing that $(w_d, z_d)$ is an invalid key pair for the cryptosystem in use.  This is a contradiction, and thus the definition of integrity is satisfied.
}

\subsubsection{Anonymity}
\label{Shuffle:proof:anonymity}

The protocol preserves integrity if
no group of $k \le N-2$ colluding members can win an {\em anonymity game},
determining which of two honest members submitted which of two plaintexts,
with non-negligible probability~\cite{
	DBLP:conf/kdd/BrickellS06}.
The attacker might gain advantage either by manipulating protocol messages,
or by using only the information revealed by a correct protocol run.
In the first case, 
the attacker can identify the intermediate ciphertext $C'_i$
of some honest member $i$
by duplicating or eliminating other honest members' ciphertexts in phase 3,
but any honest member will detect duplication in stage 3
and elimination in stage 4,
aborting the protocol before the attacker can decrypt $C'_i$.
In the second case,
an attacker who can win the anonymity game with non-negligible probability,
using only information revealed by correct protocol runs,
can use this ability to win the {\em distinguishing game}
that defines an IND-CCA2 secure cryptosystem~\cite{
	bellare1998cryptodefs,DBLP:conf/kdd/BrickellS06}.


\subsubsection{Accountability}
\label{Shuffle:proof:accountability}

A member $i$ {\em exposes} another member $j$ in phase 5b (blame)
if $i$ obtains proof of $j$'s misbehavior verifiable by a third party.
To maintain accountability,
no member may expose an honest member,
and at the end of a protocol run, either:
(a) the protocol completes successfully, or
(b) all honest members expose at least one faulty member.

We first show that no member $i$ can expose an honest member $j$.
A proof of misbehavior by $j$ consists of
some ``incriminating'' message $m_{j\phi}$ signed by $j$ in phase $\phi$,
together with all of the messages in $j$'s log up through phase $\phi$,
and the random bits each node saved during phase 2 and released in phase 5b.
Member $i$ could ``truthfully'' expose $j$
only if $j$ signs an incorrect message in phases 1--5a,
or signs more than one message per phase,
contradicting the assumption that $j$ is honest.
Member $i$ could also falsely accuse $j$
by exhibiting one of $j$'s messages $m_{j\phi}$,
together with a false ``prior'' message $m'_{k\phi'}$ (for $\phi' < \phi$)
signed by some colluding node $k$,
different from the message $m_{k\phi'}$
that $j$ actually used to compute her message $m_{j\phi}$.
In this case,
the ``proof'' will contain both $m_{k\phi'}$ (from $j$'s log)
and the false $m'_{k\phi'}$,
exposing the equivocating member $k$ instead of honest member $j$.

Now suppose a protocol run fails,
but some honest member $i$ does not expose any faulty member
in the blame phase.
Member $i$ enters the blame phase only if it:
(a) detects a faulty encryption key in phase 1,
(b) detects a duplicate or faulty ciphertext in phase 3,
(c) sees a $\goflag{j} = \false$ in phase 4,
(d) sees an incorrect $\hash{\vec{C}_N}$ in phase 4,
(e) detects a bad secondary private key in phase 5a.
Cases (a) and (e) immediately expose
the relevant message's sender as faulty.

In case (b),
member $i$ can encounter a duplicate ciphertext in phase 3
only if some member $1 \le j < i$ injected it
earlier in the anonymization phase,
or if two members $j_1$ and $j_2$ colluded to inject it in phase 2.
(Two independently encrypted ciphertexts are cryptographically unique
due to the random bits used in encryption.)
If some member $1 \le j < i$ duplicated a ciphertext,
then using the message logs of members $1$ through $i$
and the random bits from phase 2,
member $i$ can replay the decryptions and permutations
of each member before $i$ in phase 3 to expose $j$ as faulty.
If no member duplicated a ciphertext in phase 3,
then in replaying phase 3,
$i$ finds the senders of the ciphertexts $C_{j_1}$ and $C_{j_2}$
decrypting to identical ciphertexts in $\vec{C}_{i-1}$,
exposing $j_1$ and $j_2$.
If $i$ cannot decrypt a ciphertext in phase 3,
it similarly traces the bad ciphertext
to the member responsible.

In case (c) above,
either the sender $j$ of the $\goflag{j} = \false$
truthfully reported its ciphertext missing in phase 4,
or sent $\goflag{j} = \false$
although its intermediate ciphertext $C'_j$ appeared in $\vec{C}_N$.
In the former case, $i$ replays phase 3
to expose the member who replaced $j$'s ciphertext.
In the latter case, the occurrance of $C'_j$ in $\vec{C}_N$ exposes $j$ itself.

In case (d),
member $i$'s $\vec{C}_N$ does not match the $\hash{\vec{C}'_N}$
in another member $j$'s go/no-go
($\vec{C}_N \ne \vec{C}'_N$).
Members $i$ and $j$ compare message logs,
revealing that either $i$ or $j$ is lying
about the message member $N$ sent in phase 3,
or member $N$ sent two signed messages in phase 3,
exposing $i$, $j$, or $N$.


\subsection{Complexity}

If the underlying network provides efficient broadcast,
then each node transmits $O(NL)$ bits,
for a total communication cost of $O(N^2L)$.
Without efficient broadcast,
the ``normal-case'' phases 1 through 5a
still require each node to transmit only $O(NL)$ bits,
for $O(N^2L)$ overall cost,
because all broadcasts in these phases are either
single messages of length $O(NL)$
or $N$ messages of length $O(L)$.
The blame phase in an unsuccessful run
may require $O(N^3L)$ total communication
for all honest members to expose some faulty member,
but an attacker can trigger at most $O(N)$ such runs
before the group exposes and removes all faulty members.

Protocol latency is dominated by
the $N$ serial communication rounds in phase 3,
in which each node must send $O(NL)$ bits,
for a total latency of $O(N^2L)$ transmission bit-times.
Other phases require a constant number
of unicast messages or parallelizable broadcasts.

Excluding the blame phase,
each member's computational cost is dominated by
the $2N$ encryptions it must perform in phase 2,
each processing plaintexts of length $O(L+N)$
due to plaintext expansion during iterated encryption,
for an overall cost of $O(N^2+NL)$ per node or $O(N^3+N^2L)$ total.
The blame phase introduces an additional $O(N)$ factor
if all members must replay all other members' encryptions.

\com{
For a more detailed discussion of the protocol's complexity,
please see appendix \ref{appendix:complexity:Shuffle}.
}

%% file: bulk.tex

\section{Bulk Protocol}
\label{sec-bulk}

We now describe \dissent's bulk protocol formally,
prove its correctness and security,
and analyze its complexity.


\subsection{Formal Description}

Members $1, \ldots, N$ initially hold
messages $m_1,\linebreak[0]\ldots,\linebreak[0]m_N$,
now of varying lengths $L_1,\dots,L_N$.
As before,
each member $i$ has a signing key pair $(u_i, v_i)$
and a primary encryption key pair $(x_i, y_i)$;
all members know each others' public keys,
and have agreed on session identifier $n_R$
and an ordering of members.

\begin{itemize}
\item \phase{Bulk:descriptorGen} Message Descriptor Generation.
	Each member $i$ chooses a random seed $s_{ij}$ for each member $j$,
	then for each $j \ne i$,
	computes the first $L_i$ bits of a pseudo-random function
	seeded with each $s_{ij}$ to obtain ciphertext $C_{ij}$:
	\begin{center}
		$C_{ij} = \rand{L_i}{s_{ij}} ~~~ (j \ne i)$
	\end{center}
	Member $i$ now XORs her message $m_i$ with each $C_{ij}$ for $j \ne i$
	to obtain ciphertext $C_{ii}$:
	\begin{center}
		$C_{ii} = C_{i1} \oplus \ldots \oplus C_{i(i-1)} \oplus
			m_i \oplus C_{i(i+1)} \oplus \ldots \oplus C_{iN}$
	\end{center}
	Member $i$ computes $H_{ij} = \hash{C_{ij}}$,
	encrypts seed $s_{ij}$ with $j$'s public key
	to form $S_{ij} = \ER{s_{ij}}{y_j}{R_{ij}}$,
	and collects the $H_{ij}$ and $S_{ij}$ for each $j$
	into vectors $\vec{H}_i$ and $\vec{S}_i$:
	\begin{center}
		$\vec{H}_i = H_{i1}, \ldots, H_{iN}$ \\
		$\vec{S}_i = S_{i1}, \ldots, S_{iN}$
	\end{center}
	Finally, member $i$ forms a {\em message descriptor} $d_i$:
	\begin{center}
		$d_i = \{L_i, \hash{m_i}, \vec{H}_i, \vec{S}_i\}$
	\end{center}

\item \phase{Bulk:keyExchange} Message Descriptor Shuffle.
	The group runs the shuffle protocol in Section~\ref{sec-shuffle},
	each member $i$ submitting its fixed-length descriptor $d_i$.
	The shuffle protocol broadcasts all descriptors
	in some random permutation $\pi$ to all members,
	so $d_i$ appears at position $\pi(i)$ in the shuffle.

\item \phase{Bulk:dataTrans} Data transmission.
	Each member $j$ now recognizes his own descriptor $d_j$ in the shuffle,
	and sets $C'_{jj} = C_{jj}$.
	From all {\em other} descriptors $d_i$ ($i \ne j$),
	$j$ decrypts $S_{ij}$ with private key $x_j$
	to reveal seed $s_{ij}$,
	computes ciphertext $C_{ij} = \rand{L_i}{s_{ij}}$,
	and checks $\hash{C_{ij}}$ against $H_{ij}$.
	If decryption succeeds and the hashes match,
	member $j$ sets $C'_{ij} = C_{ij}$.
	If decryption of $S_{ij}$ fails or $\hash{C_{ij}} \ne H_{ij}$,
	then $j$ sets $C'_{ij}$ to an empty ciphertext, $C'_{ij} = \{\}$.

	Member $j$ now signs and sends each $C'_{ij}$
	to the designated target for the protocol run,
	in $\pi$-shuffled order:
	\begin{center}
		$\signed{j}{3}{C'_{\pi^{-1}(1)j},\dots,C'_{\pi^{-1}(N)j}}$.
	\end{center}

\item \phase{Bulk:messageRecovery} Message Recovery.
	The designated target
	(or each member if the target is the whole group)
	checks each $C'_{ij}$ it receives from member $j$
	against the corresponding $H_{ij}$ from message descriptor $d_i$.
	If $C'_{ij}$ is empty or $\hash{C'_{ij}} \ne H_{ij}$,
	then message slot $\pi(i)$ was corrupted and the target ignores it.
	For each uncorrupted slot $\pi(i)$,
	the target recovers $i$'s message by computing:
	\begin{center}
		$m_i = C'_{i1} \oplus ... \oplus C'_{iN}$
	\end{center}

\item \phase{Bulk:blame} Blame.
	If any messages were corrupted in phase \ref{Bulk:messageRecovery},
	all members run the shuffle protocol again,
	in which each member $i$ whose message was corrupted
	anonymously broadcasts an {\em accusation}
	naming the culprit member $j$:
		\begin{center}
			$A_{i} = \{j, S_{ij}, s_{ij}, R_{ij}\}$
		\end{center}
	Each accusation contains the seed $s_{ij}$ that $i$ assigned $j$
	and the random bits $i$ used to encrypt the seed.
	Each member $k$ verifies the revealed seed
	by replaying its encryption $S_{ij} = \ER{s_{ij}}{y_j}{R_{ij}}$,
	and checks that $H_{ij} = \hash{\rand{L_i}{s_{ij}}}$;
	if the accusation is valid, then each $k$ exposes $j$ as faulty.
	If the shuffle reveals no valid accusation
	for a corrupted message slot $\pi(i)$, then $k$ does nothing:
	either the anonymous sender $i$ has corrupted his own message
	or has chosen not to accuse the member who did,
	which is essentially equivalent to $i$
	sending a valid but useless message.
\end{itemize}

\subsection{Proofs of Correctness}

We now show that the bulk protocol provides
integrity, anonymity, and accountability
as defined in Section~\ref{sec-goals}.

\subsubsection{Integrity}

The shuffle protocol ensures that
the message descriptor $d_i$ of each honest member $i$
is correctly included in the shuffled output.
The target can then use either the individual ciphertext hashes $H_{ij}$ 
or the cleartext hash $\hash{m_i}$ from $d_i$
to verify the integrity of $i$'s message in the bulk output.
The cleartext hash $\hash{m_i}$ is technically redundant,
but it enables all members to verify the final results
if only one node collects and combines the ciphertexts for efficiency.

\com{
For the protocol to maintain integrity, at the end of a run of the protocol, every honest member must either (1) hold the plaintext of every other honest member, or (2) know that the plaintext of at least one honest member has been corrupted.

By way of contradiction, assume that at the end of a run of the protocol, there exists an honest member $h_{\alpha}$ who does not hold the plaintext of some other honest member and that $h_{\alpha}$ does \textit{not} know that the plaintext of an honest member $h_{\beta}$ has been modified during the protocol's execution.

The only way that honest member $h_{\beta}$'s plaintext could have been altered during the protocol's execution would be if a dishonest member sent a malformed pseduo-random string $G'_{(\pi(h_{\beta}),d)}$ in the data transmission phase, where $\pi$ is the anonymous permutation of members' message descriptors.

However, $h_{\alpha}$ would confirm in the verification phase that $H(G'_{(\pi(h_{\beta}),d)}) = H_{(\pi(h_{\beta}),d)}$.  By our assumption about the strength of the hash function, we know that the these two hash values must differ.  The fact that the hashes differ indicates to member $h_{\alpha}$ that the $h_{\beta}$'s message was corrupted during the run of the protocol.  This is a contradiction.
}

\subsubsection{Anonymity}

Suppose an attacker controls all but two honest members $i$ and $j$,
and wishes to win the anonymity game~\cite{DBLP:conf/kdd/BrickellS06}
by determining with non-negligible advantage over random guessing
which honest member sent one of their plaintexts, say $m_i$.
The attacker knows which two message slots
$\pi(i)$ and $\pi(j)$ belong to the honest members,
and must find the exact permutation $\pi$.
Since the shuffle protocol preserves anonymity
(Section~\ref{Shuffle:proof:anonymity})
and the shuffled message descriptors depend only on
random bits and the messages themselves,
the attacker learns nothing about $\pi$ from the message descriptors.
The only other information the attacker obtains about $m_i$
are the ciphertexts $C'_{ik}$ produced by all members $k$.
But since each bit of $C'_{ii}$ and $C'_{ij}$
is encrypted with a pseudo-random one-time pad
generated from a seed $s_{ij}$ that only $i$ and $j$ know,
the attacker learns nothing from these bits.


\subsubsection{Accountability}

Suppose the bulk protocol violates accountability,
implying that at the end of a protocol run,
there is some honest member $j$
who does not hold the plaintext of another honest member $i$
and does not expose any dishonest member.
Since the shuffle protocol maintains accountability,
member $j$ must have received $i$'s message descriptor $d_i$.
Since $i$ is honest,
$d_i$ contains correctly computed hashes $H_{ik}$
and correctly encrypted seeds $S_{ik}$
for ciphertexts $C'_{ik}$ that,
XORed together, would reveal $i$'s message $m_i$ to $j$.
Some member $k$
must therefore have sent an incorrect ciphertext in the bulk phase.
But since $i$ is honest,
$i$ would have sent a correct accusation of $k$ in the blame phase,
exposing $k$ as faulty.


\subsection{Complexity}

With efficient broadcast,
in the normal case each member transmits $O(N^2)$ bits
to shuffle $N$ message descriptors of length $O(N)$,
then sends $L_{tot}+O(1)$ bits in the data transmission phase,
where $L_{tot} = \sum_i L_i$.
Normal-case communication complexity is thus $O(N^2)+L_{tot}$ bits per node.
An unsuccessful run may transmit $O(N^3)+L_{tot}$ bits per node
due to the shuffle protocol's blame phase.

If $N$ is small so that message length $L_{tot}$ dominates,
if only one member wishes to transmit
($L_i=L_{tot}$ and $L_j=0$ for $j \ne i$),
and the transmitted data is incompressible,
then \dissent's communication efficiency is asymptotically optimal
for our attack model:
any member sending $o(L_{tot})$ bits cannot be the sender,
a trivial traffic analysis vulnerability.

\com{
	We assume that broadcasting a $B$-bit message to all members requires $O(B)$ bits.  The bulk protocol requires the transmission of a total of $O(N^3 + (\log{L_{max}})N^2 + L_{total}N)$ bits, where $L_{max}$ is the length of the longest message and $L_{total}$ is the sum of all message lengths.  If $N$ is small, $L_{max}$ and $L_{total}$ are very large, and $L_{max} \approx L_{total}$, then this protocol is much more efficient than the shuffle protocol.
}

The shuffle protocol incurs an $O(N^3)$ startup latency,
as the $N$ nodes serially shuffle $N$ descriptors of length $O(N)$,
but the data transmission phase is fully parallelizable,
for a total latency of $O(N^3 + L_{tot})$ transmission bit-times.

Each member $i$ performs
$N$ cryptographic operations on $O(N)$ bits each during the shuffle,
$N$ operations on $L_i$ bits to compute $C_{ii}$,
and one operation on $L_j$ bits to compute $C_{ij}$ for each $j \ne i$.
The protocol's computational complexity is thus
$O(N^2 + NL_{tot})$ per node.	

\com{
	For a more detailed discussion of the protocol's complexity, please see appendix \ref{appendix:complexity:dc}.
}

\xxx{one run's blame shuffle could be combined with the next run's
	message descriptor shuffle.}

\xxx{
	Mention variation:
	if public key cryptosystem doesn't recover randomness on decryption,
	or decryptions can fail without revealing random bits usable for replay,
	or if recovered randomness must sometimes be kept secret
	to protect the private key,
	then after a bulk protocol run in which some hash fails to match,
	just rerun the shuffle protocol
	to allow the sender of the corrupted message
	to ``accuse'' the guilty member anonymously.
	An accusation contains the random bits with which
	the sender encrypted the seed assigned to the bad $j$,
	allowing everyone to verify that $j$'s assignment was reasonable.
	If an accusation is invalid, everyone just ignores it
	and holds no one accountable;
	the sender in that slot just loses his slot
	due to his own fault.
	Perhaps put this in an appendix?
}

%% file: usage.tex
\section{Usage Considerations}
\label{sec-usage}

In describing \dissent's shuffle and bulk protocols,
we made a number of simplifying assumptions,
which we now address by placing these core protocols in the context of
a more realistic, high-level ``wrapper'' protocol.
We merely sketch this wrapper protocol without formal definition or analysis,
since it is intended only to illustrate
one way to deploy \dissent in a realistic environment,
and not to define the ``right'' way to do so.
The wrapper protocol addresses five practical issues:
protocol initiation, member selection, deniable keying,
liveness assurance, and end-to-end reliability.

\subsection{Protocol Initiation}

Our shuffle and bulk protocols assume that all group members
``just know'' when to commence a protocol run,
but in practice some node must initiate each run.
Members must {\em not} initiate a protocol run
out of a desire to send anonymously, however,
since doing so would make the sender's identity obvious to traffic analysis.

In our wrapper protocol, therefore,
each protocol run is unilaterally initiated by some node,
whom we call the {\em leader}.
To enable members to send ``spontaneously''
without compromising their anonymity,
{\em every} group member periodically initiates a protocol run
independently of its own desire to send,
on either a fixed or randomized time schedule.
(Anonymity would be equally well served
if the leader was the same for all protocol runs,
but requiring every member to act as leader occasionally
makes it easier to address liveness issues discussed below.)
If group policy permits,
a non-anonymous outsider may also lead a protocol run,
effectively invoking the collective services of the group
as in anonymous data-mining applications~\cite{
	DBLP:conf/kdd/BrickellS06}.

\subsection{Selecting Available Participants}

The core protocols above assume
that every group member participates in a given protocol run,
but in practice at least a few members of a long-lived group
are likely to be unavailable at any given time,
making it pragmatically important for the group to be able to make progress
in the absence of a few members.
The wrapper protocol therefore distinguishes
a group's {\em long-term membership} $M$
from the set of members $M_R$ participating in a particular run $R$,
where $M_R \subseteq M$.
In the wrapper protocol,
the leader of run $R$ is responsible for detecting
which members are presently available
and bringing those available to a consensus
on the precise set of participants $M_R$ for run $R$.

A key issue in choosing $M_R$
is preventing a malicious leader from packing $M_R$ with colluding members
to the exclusion of most honest members,
limiting the anonymity of the few honest members remaining.
Group policy must therefore define some minimum {\em quorum} $Q$,
and honest nodes refuse to participate in any proposed protocol run
where $|M_R| < Q$.
If there are at most $f \le Q-2$ faulty nodes, therefore,
then honest nodes are always guaranteed at least $(Q-f)$-anonymity
regardless of how $M_R$ is chosen.

As a further defense, honest members
might actively protect each other against malicious exclusion as follows.
If honest member $i$ receives a proposal from would-be leader $l_R$
to initiate run $R$ while excluding some other member $j$,
but $i$ believes $j$ to reachable,
then $i$ demands that $l_R$ add $j$ to $M_R$---%
forwarding messages between $l_R$ and $j$ if necessary---%
as a precondition on $i$ participating in round $R$ at all.

\com{
Since a few members of a long-lived group
may be be offline or unreachable at any given time,
it is usually important that the protocol be able to proceed
without the participation of some group members.
We leave it to the leader of a given protocol run to determine
which precise subset of a group's long-term membership
is connected and available to participate in a given run.
The leader initiates a run by sending all would-be participants
a signed message containing a timestamp uniquely identifying that run
and a list of the participants for that run.
If one or more nodes go offline or detectably misbehave during a protocol run
preventing the protocol from completing,
then the leader can initiate a new run
with the offline or misbehaving members excluded.
Group policy may impose requirements that the leader of any run must satisfy,
such as a {\em quorum} or minimum number of members participating in a run.
Any honest member that believes the would-be leader of a run
to be violating such a group policy,
or to be maliciously excluding members that are consistently reachable online,
can refuse to participate in a given run.
Specific policies and protocols for selecting group members
to participate in a given run
are outside the scope of this paper, however:
as a conservative approach,
the group could use byzantine fault tolerant consensus~\cite{XXX}
to make such decisions.

\xxx{note more clearly: malicious leader could try to pack one run
	with malicious members; quorum requirements protect against that.}
}

\subsection{Coercion Resistance via Deniable Keying}
\label{sec-coercion}

\Dissent's shuffle protocol assumes each group member $i$ has
and a signing key pair $(u_i,v_i)$ with which it signs all messages,
creating the nonrepudiable ``accountability trail''
that the blame phase (5b) requires to trace a misbehaving member.
Unfortunately, this nonrepudiable record could also enable members
to prove to a third party which message they sent (or didn't send)
in a given protocol run.
In anonymous communication scenarios
we often desire not just anonymity but also repudiability~\cite{
	borisov04offtherecord}:
after a protocol run, no one should be able to prove to a third party
which message any member sent, or ideally,
whether a member participated at all.
In anonymous voting applications,
we often desire the closely related property
of resistance to coercion or ``vote-buying.''

Our wrapper protocol can provide
a form of repudiability or coercion resistance as follows.
We assume each group member $i$'s well-known identity
is defined {\em only} by its primary encryption key pair $(x_i,y_i)$,
and members now choose a fresh, temporary signing key pair $(u_i,v_i)$
for each protocol run.
To initiate a run,
the would-be leader $l$
uses a deniable authenticated key exchange algorithm such as SKEME~\cite{
	raimondo05secure}
to form a secure channel with each potential participant $i$,
using $l$'s and $i$'s primary encryption keys for authentication.
Each member $i$ uses this pairwise-authenticated channel
to send the leader $i$'s fresh public signing key $v_i$ for the run.

Once $l$ forms a tentative list of $N = |M_R|$ participants,
$l$ broadcasts to all participants
a {\em round descriptor} $D_R$ containing
a timestamp, all participants' primary public keys $y_1,\dots,y_N$,
and all participants' corresponding
temporary signing keys $v_1,\dots,v_N$ for the run.
Each member $i$ now forms a {\em challenge} $c_{ij}$ for each node $j$,
containing a random nonce $N_{ij}$ and a hash of $D_R$ keyed on $N_{ij}$.
Member $i$ then encrypts $c_{ij}$ with $j$'s public key $y_j$ to yield $C_{ij}$.
Member $i$ sends its encrypted challenges to the leader,
who forwards each $C_{ij}$ to member $j$.
Member $j$ decrypts $C_{ij}$,
verifies the keyed hash it contains against the $D_R$
that $j$ received from the leader,
and returns $c_{ij}$ to the leader, who forwards it to $i$.
On a decryption failure or challenge mismatch,
the leader must decide whether to exclude $i$ or $j$ from a retry attempt;
$i$ can prove its innocence by revealing the random bits it used
to encrypt its original challenge to $j$.

Once all members confirm $D_R$ with all other members,
the shuffle protocol proceeds using the temporary signing keys in $D_R$.
These signing keys provide nonrepudiation only {\em within the protocol run},
allowing the leader to trace misbehaving members
and exclude them from subsequent runs.
No node is left with proof that any member $i$
actually used signing key $y_i$ during a given run, however,
since anyone can unilaterally forge all the authenticated key exchanges,
challenges, and subsequent messages in the shuffle and bulk protocols.

Of course, this form of repudiability is useful
only against an attacker who actually requires third-party verifiable
``proof of responsibility''
in order to coerce group members.
If the attacker can see all network traffic, as our attack model assumes,
{\em and} the attacker's traffic logs alone constitute ``proof''
of which network packets a given member sent,
then we know of no way to achieve deniability or coercion resistance.
Similarly,
a member might be coerced {\em before} a protocol run
into sending some sufficiently unique, attacker-supplied message or ballot;
if the mere appearance of that message/ballot in the run's output
satisfies the attacker that the member ``stayed bought,''
then no anonymity mechanism based purely on a random shuffle
will address this form of coercion.

\subsection{Ensuring Liveness}

As we have seen,
tracing active disruptors of the shuffle or bulk protocols
presents particular technical challenges
due to the need to protect the anonymity of honest senders.
A member might {\em passively} disrupt either protocol, however,
by simply going offline at any time,
either intentionally or due to node or network failure.
Fortunately, given the core protocols' resistance
to both active disruption and traffic analysis,
we can ensure liveness and handle passive disruption
via more generic techniques.

Each phase of the shuffle and bulk protocols
demand that particular members send properly signed messages
to other members.
Again borrowing terminology and ideas from PeerReview~\cite{
	haeberlen07peerreview},
when the protocol demands that member $i$ send member $j$ a message,
and member $j$ has not received such a (properly signed) message
for some time,
we say that $j$ {\em suspects} $i$.
Once $j$ suspects $i$, $j$ informs another node $k$ (the leader, for example)
of $j$'s suspicion;
$k$ in turn contacts $i$ demanding a (signed) copy of $i$'s message to $j$.
If $i$ fails to offer this message to $k$,
then after some time $k$ suspects $j$ as well
and notifies other members in turn,
eventually causing all honest, connected members to suspect $i$.
Member $i$ can dispel any honest member's suspicion at any time
by offering a copy of the demanded message.
If $i$ honestly cannot send to $j$ due to asymmetric connectivity, for example,
then $i$ responds to $k$'s demand with the required message,
which $k$ forwards back to $j$,
dispelling both $j$'s and $k$'s suspicion and enabling the protocol to proceed.

Since our wrapper protocol makes
the leader responsible for initiating protocol runs,
we also make it the leader's responsibility
to decide when a protocol run has failed
due to a suspected node going offline---%
or deliberately withholding a required message---%
for too long.
At this point,
the leader starts a new protocol run,
excluding any exposed or persistently suspected nodes from the previous run,
and the remaining members attempt to resend their messages.
If the leader fails,
members can retry their sends in a future run initiated by a different leader.


\subsection{End-to-End Reliability}

A corner-case liveness challenge for most protocols
is {\em closure}, or determining when participants
may consider the protocol ``successfully concluded.''
In a byzantine model,
a malicious member
might intentionally withhold the last message he was supposed to send---%
e.g., his own secondary private key in phase 5a of the shuffle protocol,
or his own ciphertext in the bulk protocol---%
while collecting the last messages of other members,
thereby learning the results of the protocol run
while denying those results to other members.

We approach this class of problems in general
by treating our shuffle and bulk protocols
as a ``best-effort'' anonymous delivery substrate,
atop which some higher-level protocol must
provide end-to-end reliable delivery and graceful closure if desired.
If a faulty member denies other members a protocol run's results,
the honest members will soon suspect the faulty member,
and the same or a different leader will eventually start a new protocol run
without the faulty member,
in which the members may retransmit their messages.
If a member $i$ wishes to ensure that a message he sends anonymously
is reliably seen by a particular member $j$,
$i$ must resend the message in successive protocol runs
until $j$ acknowledges the message.
(Member $j$ might sign acknowledgments
via either a public or pseudonymous key).

If the messages sent in a protocol run are interrelated,
such as the ballots comprising an anonymous vote,
and the group wishes to ensure that some quorum of members see the result,
then the group can follow the voting run with an acknowledgment run,
discarding and repeating unsuccessful votes
(with successively smaller membership sets if members go offline)
until the required number of members acknowledge the results.
If the group wishes to provide reliable broadcast semantics
or maintain some consistent group state across successive protocol runs,
the group can implement byzantine consensus~\cite{castro99practical}
atop the shuffled send primitive,
ensuring both liveness and strong consistency
as long as over two thirds of the group members remain live.

%% file: impl.tex
\section{Prototype Implementation}
\label{sec-impl}


To evaluate \dissent's practicality,
we built and tested a simple prototype of the protocol.
The prototype is written in Python,
using OpenSSL's implementations of
1024-bit RSA-OAEP with AES-256 for public-key encryption and signing,
AES-256 in counter mode as the bulk protocol's pseudo-random function,
and SHA-1 as the hash algorithm.
\com{
It took only 3--4 days to get an initial version of the protocol running,
followed by a week to work out performance issues;
we have done no extensive optimization to the prototype, however.
}

We used the Emulab~\cite{emulab} network testbed 
to test the prototype under controlled network conditions.
We ran the prototype on recent x86 PCs machines
running Ubuntu 7.04 and Python 2.5,
on a simulated star topology
in which every node is connected to a central switch 
via a 5Mbps connection with a latency of 50ms
(100 ms node-to-node latency).
We make no claim that this topology is ``representative''
of likely deployment scenarios for \dissent,
since we know of no data on the network properties
``typical'' online groups that might wish to run \dissent.
Our simulated topology is merely intended to reflect {\em plausible}
communication bandwidths and delays
for wide-area Internet communication.

We rely on the formal analysis in previous sections
to evaluate \dissent's security properties,
and assume that the accountability measures
in a full implementation of \dissent
will deter or eventually exclude misbehaving members.
For experimentation purposes, therefore,
we implement and test only the ``normal-case'' aspects of the protocol
in the current prototype.
The prototype does not use a secure public key infrastructure,
and does not implement the ``blame'' phases
or the wrapper protocol.
Nodes sign and verify all messages, however,
ensuring that performance measurements accurately reflect
\dissent's normal-case costs.

\com{
We do not use a public key infrastructure, but instead
have every node broadcast both its 
secondary (one-time use) \emph{and} primary (long-term)
public keys at the start of the protocol.  This method 
of key exchange is entirely insecure, but it
is suitable for performance-testing purposes.
}

The prototype uses TCP for communication,
maintaining TCP connections throughout a given protocol run
to minimize startup overhead,
but closing all connections at the end of each run.
Where \dissent requires broadcast,
nodes implement these broadcasts atop TCP
by sending their messages to a leader,
who bundles all broadcasts for that phase
and sends each node a copy of the bundle.

\com{	not critical
We begin a protocol run by instructing a control
node to invoke a Python script on each test machine.
The control node directs each node to execute
either the shuffle protocol alone,
or the shuffle and bulk protocols together.
The control node also seeds each node 
with per-round information such as round identifier,
node identifier, message length, and other nodes' addresses.
}

\subsection{Performance Evaluation}

\begin{figure}[t]
\centering
\includegraphics[width=0.45\textwidth]{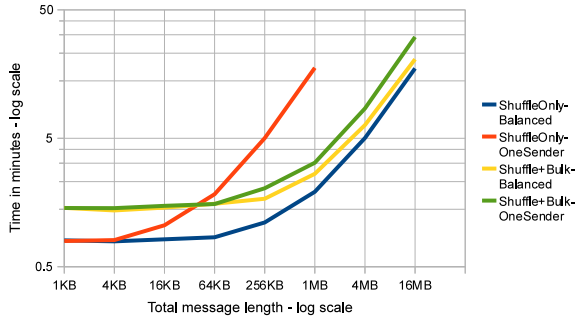}
\caption{Time required for anonymous broadcast
	of balanced and unbalanced message loads among 16 nodes,
	via shuffle alone or full \dissent protocol.}
\label{fig-results-varlen}
\end{figure}

Figure \ref{fig-results-varlen} shows the total time
the prototype requires to broadcast messages of varying sizes anonymously
among $16$ nodes,
using either the shuffle protocol alone or the full \dissent protocol.
In each case,
we test two message loads:
a {\em Balanced} load in which each node sends $1/16$th
of the total message data,
and a {\em OneSender} load in which one node sends all the data
and other nodes send nothing.

For a single node to send a $16$MB message, \dissent
runs in about 31 minutes, or $3.6\times$
longer than one node requires to broadcast the same data to the other 15 nodes
with no encryption or anonymization.  
While significant, we feel this is a reasonable price to pay
for strong anonymity.
\com{	would have to explain and back up this ``theoretical minimum'' claim...
That is $1.7$ times longer than the theoretical minimum amount
of time it would take a node to anonymously broadcast
a $16$MB message across our testbed network.
}

As expected,
the full protocol incurs a higher startup delay
than the shuffle protocol alone,
but handles unbalanced loads more gracefully,
maintaining similar performance for a given total message length
regardless of balance.
We are not aware of any other verifiable shuffles~\cite{
	neff2001voting,furukawa01efficient}
for which working implementations and performance data are available,
but given their typical assumption of small, equal-length messages,
we expect their performance on unbalanced loads
to be at best on par with our shuffle protocol alone.

\com{
The performance bottleneck for both protocols was network communication,
not computation.
As a result, the full \dissent protocol
overwhelmingly outperformed the shuffle-only protocol
on unbalanced message loads.

Maintaining anonymity in the shuffle-only protocol requires
each group member to send a message of equal length.  When 
message lengths vary, the shuffle-only protocol
requires all members to pad their messages up
to the length of the longest message.  The deadweight loss
of transmitting these padding bits over the network is
substantial when the variance in message length is large.
\Dissent requires no such padding, so the \dissent protocol
outperforms the shuffle-only protocol in these cases.

When every group member sends a message of \emph{exactly} the same
length, the shuffle-only version of the protocol runs slightly
faster than \dissent.  In the equal-length case,
the shuffle-only protocol requires transmission of no padding bits.
The overhead incurred by \dissent{}'s message descriptor
shuffle phase makes \dissent slightly slower in these
cases.
}

\begin{figure}[t]
\centering
\includegraphics[width=0.45\textwidth]{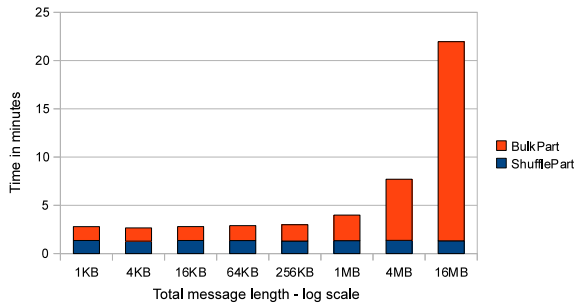}
\caption{Time required to send varying message sizes,
	broken into shuffle and bulk transfer protocol portions.}
\label{fig-results-parts}
\end{figure}

Figure \ref{fig-results-parts} 
breaks the runtime of the full \dissent protocol
into its shuffle and bulk protocol components,
illustrating that the shuffle's cost remains constant with message size
and becomes negligible as total message length grows.

The full \dissent protocol still shows some slowdown
under highly unbalanced load:
although balance does not affect \dissent's communication cost,
it {\em does} affect computation costs.
When only one node is sending,
that node must compute and XOR together $N-1$ pseudo-random streams
of message length $L$,
while other nodes each compute only one $L$-byte stream.
This timing difference could lead to a side-channel attack
if not handled carefully in implementation,
e.g., by pre-computing all required bit strings
before commencing a send.
We have made no attempt to analyze the protocol in detail
for side-channel attacks, however.

\com{
Over the course of a run of the bulk
transfer phase, all $N$ nodes generate a total of
$2(N-1)L_{total}$ pseudo-random bits.  Each pseudo-random
bitstring is generated twice: the author of a given
message slot XORs $N-1$ pseudo-random strings with the
data message, and $N-1$ non-authoring participants generate
$1$ pseudo-random string for each message slot.

When all members transmit equal length messages, the 
computational cost of generating the $2(N-1)L_{total}$
psuedo-random bits is distributed among all $N$ group members.
If only one member sends data, then that authoring
node bears the entire computational cost of generating
$(N-1)L_{total}$ pseduo-random bits.  If nodes have equal
computational power, the message-descriptor
generation phase will take slightly longer in this case
than in the equal-length case.  

Note that
in a deployment of \dissent, the amount of time a node
takes to generate a message descriptor could reveal
information about the length of that node's message.
An implementation could avoid this vulnerability
by having nodes generate
their message descriptors before the start of a protocol
run.  We have not investigated opportunities for more
sophisticated timing attacks against \dissent, but
a deployment of the protocol would have to take these
into account.
}

\com{	Ignore logarithmic factors for simplicity -
	crypto folks generally assume the security parameter is a constant,
	and the security parameter can be no smaller than the log 2
	of message length...  -baf }

\begin{figure}[t]
\centering
\includegraphics[width=0.45\textwidth]{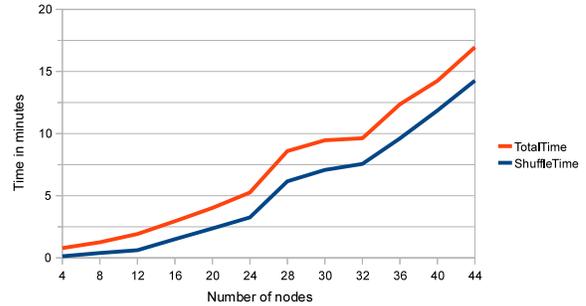}
\caption{Time required to send 1MB of data (balanced) using
	shuffle and bulk transfer protocols together
		with a varying group size.}
\label{fig-results-varnodes}
\end{figure}

Figure \ref{fig-results-varnodes} measures the prototype's runtime
with varying group sizes.
In a successful run,
each node sends $O(N^2)$ bits in the shuffle
and $L_{tot} + O(1)$ bits in the bulk protocol.
As expected,
the shuffle's runtime increases much more quickly with $N$
than the bulk protocol,
although the superlinear $N^2$ curve manifests only slightly
for the small groups we tested.

%% file: related.tex
\section{Related Work}
\label{sec-related}

\com{
This paper defines \dissent, a new \emph{shuffled send} primitive
that is computationally anonymous, maintains message integrity,
enforces a one-to-one mapping of group members to 
messages, and prevents attacks against protocol progression
(DoS attacks).
Previously proposed protocols offer only some, 
not all, of the desirable properties that our protocol achieves. 
}

\dissent's shuffle protocol builds directly on
Brickell and Shmatikov's data collection protocol~\cite{
	DBLP:conf/kdd/BrickellS06},
adding DoS resistance via our new go/no-go and blame phases.
\dissent's bulk protocol is similarly inspired by DC-nets~\cite{
	chaum88dining},
which are a computationally efficient and provide unconditional anonymity,
but traditionally require nondeterministic ``reservation'' schemes
to allocate the anonymous channel's communication bandwidth,
and are difficult to protect against anonymous DoS attacks by group members.
Strategies exist to strengthen DC-nets
against DoS attacks~\cite{waidner89dining},
or to form new groups when an attack is detected~\cite{sirer04eluding}.
\Dissent's use of a shuffle protocol
to set up a {\em deterministic} DC-nets instance, however,
cleanly avoids these DoS vulnerabilities
while providing the additional guarantee
that each member sends {\em exactly} one message per protocol run,
a useful property for holding votes or assigning 1-to-1 pseudonyms.

Mix-networks~\cite{chaum81untraceable} provide
scalable and practical anonymous unicast communication,
and can be adapted to group broadcast~\cite{perng2006multicasting}.
Unfortunately,
mix-networks are difficult to protect against traffic analysis~\cite{
	serjantov2003mixattacks}
and DoS attacks~\cite{
	dingledine02reliable,iwanik04duo},
and in fact lose security under DoS attack~\cite{
	borisov07denial}.
\com{By timing the movement of packets among the nodes
of a mix network, an attacker can make reasonably good
guesses about the identity of the sender and receiver
of encrypted packets traveling through the network.
Given enough traffic flow data, the attacker completely
compromises the anonymity of the group members.
Our protocol is not susceptible to this type of statistical
timing attack or to traffic analysis.}
Crowds \cite{reiter1999crowds}
are more computationally efficient that mix networks,
but are vulnerable to statistical traffic analysis
when an attacker can monitor many points across
the network.
\Dissent in contrast is provably secure against traffic analysis.

Anonymous voting protocols solve a problem that closely relates
to the group broadcast problem.  Each user casts a ballot
whose contents should be publicly known but whose author should
be unknown to both the election officials and other voters.
Many voting protocols allow 
transmission of only fixed-length ``Yes'' or ``No'' messages 
\cite{adida2006voting}.

Cryptographically verifiable shuffles~\cite{
	neff2001voting,furukawa01efficient}
might be used in place of our shuffle protocol,
allowing shuffles to be performed and verified offline.
These algorithms require more complex calculations, however.
Further, guaranteeing not only a shuffle's correctness,
but also its {\em randomness} and hence anonymity
in the presence of compromised members,
still requires passing a batch of messages
through a series of independent shuffles,
as in \dissent or mix-networks~\cite{dingledine04synchronous}.
\com{ other verifiable shuffles and mix-type voting schemes:
	Sako, Receipt-free mix-type voting scheme, Eurocrypt'95
	Abe, Mix-Networks on Permutation Networks - ASIACRYPT 99,
	Abe and Hoshino. Remarks on Mix-Network Based on ..., PKC'01
	An Efficient Scheme for Proving a Shuffle, CRYPTO'01
	Neff, A Verifiable Secret Shuffle, CCS'01
	Groth, A Verifiable Secret Shuffle of Homomorphic Encryptions '03

Also relevant to the cascade requirement:
	Dingledine/Shmatikov/Syverson,
	"Synchronous Batching: From Cascades to Free Routes"
}

Other relevant schemes for group-oriented anonymity include
ring signatures \cite{rivest2001ring}, which provide no 
protection against traffic analysis, and $k$-anonymous
message transmission protocols \cite{ahn2003kanonymous}, which provide
anonymity only when a large fraction of group members
are honest.

Tor~\cite{dingledine2004tor} and Herbivore~\cite{sirer04eluding}
are two well-known practical systems for providing anonymous communication
over the Internet.  These systems scale to far larger
groups than \dissent does, and
also permit interactive communication.
These systems do not provide \dissent's strong guarantees
of anonymity or accountability, however.
As a system based on mix networks,
Tor is vulnerable to traffic analysis attacks.
Herbivore provides unconditional anonymity,
but only within a small subgroup of the total group of participants.
\Dissent may be more suitable
for non-interactive communication between
participants willing to sacrifice protocol execution speed
for strong assurances of anonymity and accountability.

%% file: concl.tex
\section{Conclusion}
\label{sec-concl}

\Dissent is a novel protocol for 
anonymous and accountable
group communication.
\Dissent allows a well-known
group of participants to anonymously exchange variable-length
messages without the risks of traffic analysis or anonymous
DoS attacks associated with mix-networks and DC-nets.
\Dissent improves upon previous shuffled-send primitives
by adding accountability -- the ability to trace faulty
nodes -- and by eliminating the
message padding requirements that limit
earlier schemes.
We have reviewed the practical concerns associated
with a real-world deployment of \dissent, and we have proposed
possible solutions for each.
Our implementation demonstrates that \dissent
is a practicable protocol, at least
for a medium-sized group of participants.

\com{	XXX include these in the final paper, if/when we get there.
\subsection*{Acknowledgments}
\xxx{Add these}
Vitaly, Mike Fischer, MPI-SWS students, ...
Justin Brickell, Jacob Strauss, Pedro Fonseca,
Funding
}

%% file: complexity.tex
\section{Detailed Complexity Analysis}

\subsection{Shuffle Protocol}	

	\label{appendix:complexity:Shuffle}
	\subsubsection{Computation}
	
	Computationally expensive operations include keypair
	generation, encryption, decryption, message signing, 
	hashing, and signature verification.  In phase \ref{Shuffle:keypairGen}, each participant generates a secondary keypair, signs the keypair, and verifies the $N-1$ signatures on the other participants' keypairs.  In phase \ref{Shuffle:dataSubmission}, each participant performs $2N$ encryptions and signs 1 message.  In phase \ref{Shuffle:anonymization}, participant 1 verifies $N-1$ signatures and other participants verify $1$ signature.  In phase \ref{Shuffle:anonymization}, all participants also perform $N$ decryptions and sign $1$ message.  In phase \ref{Shuffle:verification}, each participant verifies $1$ signature, from participant $N$, computes 1 hash, and signs one ``Go" message.  After receiving the ``Go" messages in phase \ref{Shuffle:verification}, every participant verifies the $N-1$ signatures on these messages.
	
	If participants enter the ``blame" phase (phase \ref{Shuffle:blame}b), participant 1 verifies $N-1$ signatures.  Every other participant must verify the $N-1$ signatures on the original ciphertexts $C'$, and $N-1$ signatures on the permuted data -- a total of $2N-2$ verifications.  Every participant also encrypts $N$ ciphertexts $N$ times to confirm the honesty of all other participants.  This is $N^2$ encryptions.
	
	Otherwise, in the decryption phase (phase \ref{Shuffle:decryption}a), each participant verifies $N-1$ signatures and performs $N$ decryptions for each of $N$ ciphertexts for a total of $N^2$ decryptions.
	
	\subsubsection{Communication Rounds}
	
	Phase \ref{Shuffle:keypairGen} requires 1 parallelizable broadcast round for participant to exchange secondary public keys.  Phase \ref{Shuffle:dataSubmission} requires 1 round of communication for each participant to submit her data to participant 1.  Phase \ref{Shuffle:anonymization} requires $N-1$ rounds -- one for each pair of participants between 1 and $N$.  Phase \ref{Shuffle:verification} requires 1 round for participant 1 to broadcast the set of ciphertexts and 1 parallelizable broadcast round for each participant to broadcast her ``Go" message.
	
	Phase \ref{Shuffle:decryption}a requires 1 parallelizable broadcast round for each participant to broadcast her secondary private key.  Phase \ref{Shuffle:blame}b requires 1 parallelizable broadcast round for each participant to submit to all other participants her proof-of-correctness data.
	
	The protocol requires a total of 4 parallelizable broadcast rounds and $N+1$ simple rounds.  In the worst case, each parallelizable broadcast round can be implemented with $2$ simple communication rounds -- each participant sends their broadcast message to participant 1, who concatenates each and sends the string of broadcast messages to each participant.  Using this implementation results in a total of $N+9$ serial communication rounds, or $O(N)$ rounds.
	
	\subsubsection{Total Data Exchanged}
	
	Denote the length of a data message as $L$, the length of a key as $T$, the length of a signature as $Q$, the length of a hash as $H$, and denote the length of the randomness used for a single public-key encryption as $R$.  Since the round identifier $n_R$ is sent with every signature, we include the length of $n_R$ in $Q$.
	
	In phase \ref{Shuffle:keypairGen}, participants broadcast their signed public keys -- a total of $(T+Q)N$ bits broadcasted among all participants.  In phase \ref{Shuffle:dataSubmission}, all $N$ participants send their signed ciphertext to participant 1, for a total of $(L+Q)N$ bits.  In phase \ref{Shuffle:anonymization}, each participant sends a message of length $LN+Q$.  The total data exchanged in phase \ref{Shuffle:anonymization} is then $LN^2+QN$.
	
	In phase \ref{Shuffle:verification}, participant 1 broadcasts a message of length $LN+Q+1$.  Other participants broadcast messages of length $H+Q+1$.  In phase \ref{Shuffle:verification}, there are $(L+Q+H)N+Q+1$ bits broadcasted.  
	
	Participants run either phase \ref{Shuffle:decryption}a or \ref{Shuffle:blame}b.  In phase \ref{Shuffle:decryption}a, each participant broadcasts her private key, for a total of $(T+Q)N$ bits broadcasted.  If they enter phase \ref{Shuffle:blame}b, participant 1 broadcasts a message of length $(L+R+Q)N+Q$ and other participants broadcast messages of length $(2L+R)N+2Q$.  The total data exchanged in phase \ref{Shuffle:blame}b is then $(2L+R)N^2+(L+R+3Q)N+Q$.

	As the number of participants in $N$ gets large,
	the amount of data exchange will become dominated by 
	the \linebreak[4] $(2L+R)N^2$ term in the blame phase and the $LN^2$ 
	term in the anonymizaton phase. Since $R$ will generally 
	be either constant or proportional in length to $L$, we
	can simplify the bound to $O(LN^2)$.
	
	We assume that a $B$-bit broadcast message can be sent to all $N$ participants in $O(B)$ bits.  If a $B$-bit broadcast message requires $O(BN)$ bits to transmit, then the upper bound on the number of bits transmitted must be increased to $O(LN^3)$.  This is because, in the blame phase, participants broadcast $N$ messages of length $O(LN)$.  Outside of the blame phase, participants in total broadcast either $N$ messages of $O(L)$ bits or one participant transmits a message of $O(LN)$ bits, so the complexity for all other phases is still $O(LN^2)$ when using an inefficient broadcast method.

\subsection{Bulk Protocol}
	\label{appendix:complexity:dc}
	\subsubsection{Computation}
	
	Computationally expensive operations include \linebreak[2] encryption, 
	decryption, message signing, signature verification, 
	hashing, pseudo-random string generation, and cryptographically
	secure random number generation.
	
	In phase \ref{Bulk:descriptorGen}, generates $N-1$ pseudo-random strings, calculates $N$ message hashes and performs $N$ encryptions -- $3N-1$ operations per participant.
	
	In phase \ref{Bulk:keyExchange}, participants run the shuffle protocol, with an overall computational complexity per node (as demonstrated above) of $O(N^2)$.  Each participant then decrypts the $N$ seeds exchanged in the protocol.
	
	In phase \ref{Bulk:dataTrans}, each participant generates $N-1$ pseudo-random strings and calculates $N-1$ hashes.  In phase \ref{Bulk:verification}, each participant computes at most $N^2$ hashes and $N^{2}$ encryptions for a total of $2N^2$ computations.
	
	The dominant computational costs of this algorithm are the anonymous key exchange and the pseudo-random string verification in phase \ref{Bulk:verification}.  The algorithm requires $O(N^2)$ computational operations per participant for $O(N^3)$ computations overall.
	
	\subsubsection{Communication Rounds}
	
	Phase \ref{Bulk:descriptorGen} requires no communication.  Phase \ref{Bulk:keyExchange} uses the shuffle protocol above and therefore requires $O(N)$ communication rounds.  Phase \ref{Bulk:dataTrans} requires one parallelizable broadcast round.  Phase \ref{Bulk:verification} and \ref{Bulk:messageRecovery} require no communication.  
	
	If parallelizable broadcast rounds are implemented by having each node send their broadcast to a ``leader" node and the leader relaying all messages to each node, then each parallelizable broadcast round requires 2 communication rounds.  The total number of communication rounds is then $O(N)$.

	\subsubsection{Total Data Exchanged}

	We assume that broadcasting a $B$-bit message to
	all participants requires $O(B)$ bits.  Denote the length
	of the \linebreak[4] maximum-length data message as $L_{max}$, the sum
	of message lengths as $L_{total}$, the length of a signature
	as $Q$, the length of a hash as $H$, the length of the
	randomness used to encrypt the seed as $R$, and the 
	length of a pseudo-random number generator seed as $P$.
	Since the round identifier $n_R$ is always transmitted
	with the signature, we include the length of $n_R$ in 
	the length of a signature $Q$.

	In phase \ref{Bulk:descriptorGen}, no data is transmitted.  In phase \ref{Bulk:keyExchange}, participants run the shuffle protocol using messages of length $\log_2{L_{max}} + (P+H)N$.  The shuffle protocol requires transmission of $O(LN^2)$ bits for an $L$-bit message, so the total data exchange complexity of phase \ref{Bulk:keyExchange} is $O(((P+H)N + \log{L_{max}})N^2)$.  Since $P$ and $H$ are constants of the cryptosystem in use, we simplify this to $O(N^3 + (\log{L_{max}})N^2)$ bits.
	
	In phase \ref{Bulk:dataTrans}, each participant transmits one pseudo-random string for each message.  Each participant broadcasts a message of length at most $L_{total}+(P+R+1)N+Q$, or $O(L_{total} + N)$ bits.  All participants together then transmit $O(N^2 + NL_{total})$ bits.
	
	Phases \ref{Bulk:verification} and \ref{Bulk:messageRecovery} require no data exchange.
	
	The amount of data exchanged is dominated by the cost of the anonymous data exchange protocol in phase \ref{Bulk:keyExchange}.  The overall data exchange complexity is $O(N^3 + (\log{L_{max}})N^2 + L_{total}N)$.  If we do not assume that broadcasts are efficient, then this bound increases to $O(N^4 + (\log{L_{max}})N^2 + L_{total}N)$ bits.